\documentclass[11pt,a4paper]{article}
\usepackage{jheppub}

\usepackage{amsmath,amssymb,graphicx}
\usepackage{xcolor}
\usepackage{tikz}
\usepackage{orcidlink}
\usepackage{hyperref}
\newcommand{\rs}{r_s}
\newcommand{\AH}{A_{\rm H}}
\newcommand{\Thaw}{T_{\rm H}}
\newcommand{\ph}{\phi}
\newcommand{\dd}{\mathrm{d}}

\begin{document}

\title{The Fluctuation--Dissipation Theorem and Dynamical Love Number
in Schwarzschild Black Holes}

\author[a,b]{D.~Giataganas\orcidlink{0000-0003-2003-3902},}
\author[c]{A.~Kehagias\orcidlink{0000-0001-6080-6215},}
\author[d]{A.~Riotto\orcidlink{0000-0001-6948-0856}}

\affiliation[a]{Department of Physics, National Sun Yat-sen University, Kaohsiung 80424, Taiwan
}
\affiliation[b]{Physics Division, National Center for Theoretical Sciences, Taipei 10617, Taiwan}
\affiliation[c]{Physics Division, National Technical University of Athens, Athens 15780, Greece}
\affiliation[d]{Department of Theoretical Physics and Gravitational Wave Science Center,  \\
Universit\'e de Gen\`eve, 24 quai E. Ansermet, CH-1211 Geneva 4, Switzerland}

\abstract{
We study the real-time response of a Schwarzschild black hole in the Hartle--Hawking state to an external massless scalar field. Using a complexified radial contour to represent the Schwinger--Keldysh closed time path, we describe dissipation, thermal fluctuations and conservative tidal response within a common framework. Before performing tidal matching, we derive the equilibrium fluctuation--dissipation relation directly from the radial construction for every scalar multipole at arbitrary real frequency. In the low-frequency tidal expansion, we recover the known relation between the logarithmic running of the conservative dynamical Love coefficient at order $\omega^2$ and the leading dissipative response at order $\omega$. We explain how the same leading absorptive coefficient enters both relations at low frequency and trace the unit coefficient of the running relation to the Schwarzschild horizon slope. In the Hartle--Hawking state, this common coefficient relates the logarithmic running directly to the zero-frequency thermal noise. Kubo--Martin--Schwinger analyticity determines the thermal noise from the dissipative response, whereas the near-zone radial equation and tidal matching determine the conservative logarithmic running. 
}

\emailAdd{dimitrios.giataganas@gmail.com}
\emailAdd{kehagias@central.ntua.gr}
\emailAdd{Antonio.Riotto@unige.ch}
\maketitle
\flushbottom

\section{Introduction}
\label{sec:intro}
\noindent
A compact object placed in a slowly varying external field develops an induced multipole moment. Its linear response contains a conservative part, described at low frequency by static and dynamical Love coefficients, and a dissipative part associated with absorption \cite{Flanagan:2007ix,Rodriguez:2026iot,Chakraborty:2026qru}. The conservative static Love numbers vanish for four-dimensional Schwarzschild and Kerr black holes \cite{Binnington:2009bb,Damour:2009vw,Charalambous:2021mea,Kehagias:2024rtz,Combaluzier:2025eoc,Gounis:2024hcm}. For the Schwarzschild scalar problem considered here, the leading nonvanishing response is absorptive and linear in frequency, while the leading conservative dynamical response appears at order $\omega^2$ and contains logarithmic scale dependence. Causality and positivity arguments provide generic bounds on the tidal Love numbers of gravitating objects \cite{Kehagias:2026opd,Apostolidis:2026mvb,Kehagias:2026nqi}.

In the worldline effective theory of compact objects, the conservative response is encoded in local worldline operators, while dissipation is described by correlation functions of internal worldline degrees of freedom. Their coefficients are fixed by matching to the full gravitational solution \cite{Goldberger:2005cd,Goldberger:2020fot,Goldberger:2020wbx,Rodriguez:2026iot}.

Recent calculations have revealed a simple relation between the leading dissipative response and the logarithmic running of the leading dynamical Love coefficient of Schwarzschild black holes~\cite{Charalambous:2021mea,Perry:2023wmm,Ivanov:2024sds,Combaluzier:2025eoc}. In the conventions used here,
\begin{equation}
\omega\rs\,{\rm Im}\,K_\ell(\omega) = -\mu\frac{\dd}{\dd\mu}K_\ell(\omega) +{\cal O}(\omega^3),
\label{eq:645}
\end{equation}
where $K_\ell$ is the retarded tidal response kernel, $\mu$ is the
renormalization scale, and $\rs$ is the Schwarzschild radius;  see Ref.~\cite{Rodriguez:2026iot} for a review. At the order displayed, the left-hand side is determined by the leading horizon absorption, while the right-hand side extracts the logarithmic running of the conservative response. Dimensional analysis in systems with a unique scale allows their coefficients to differ by an arbitrary dimensionless number, yet for Schwarzschild that number is unity~\cite{Combaluzier:2025eoc}. Related structures have been found at higher orders in the frequency expansion, where the most transcendental terms continue to display a close connection between conservative and dissipative response, although the relations become more intricate \cite{Caron-Huot:2025tlq,Combaluzier:2025eoc}. This has motivated the interpretation of \eqref{eq:645} as a fluctuation--dissipation type correspondence \cite{Combaluzier:2025eoc,Rodriguez:2026iot}. As already noted in Ref.~\cite{Combaluzier:2025eoc}, however, \eqref{eq:645} involves only the causal response.

The genuine fluctuation--dissipation theorem concerns a different relation. In a thermal state it connects the dissipative part of the retarded kernel to the symmetric, or noise, correlator,
\begin{equation}
F_\ell(\omega) = \coth\!\left(\frac{\beta\omega}{2}\right){\rm Im}\,K_\ell(\omega),
\label{eq:fdtintro}
\end{equation}
where $\beta$ is the inverse temperature. Unlike \eqref{eq:645}, this relation follows from the Kubo--Martin--Schwinger (KMS) condition~\cite{Kubo:1957mj,Martin:1959jp} and is naturally implemented in a real-time formalism that keeps track of both response and fluctuations.

One purpose of this paper is to derive \eqref{eq:fdtintro} directly for a Schwarzschild black hole and to determine precisely how it is related to, and distinguished from, the running relation \eqref{eq:645}. To do so, we use the Schwinger--Keldysh closed-time-path formalism to describe the causal response and equilibrium fluctuations within a single generating functional
\cite{Schwinger:1960qe,Keldysh:1964ud,Chou:1984es,Kamenev:2011ba}.

In the standard gravitational realization of this formalism, the two branches of the closed time path are represented by two Lorentzian segments joined to a Euclidean segment that prepares the thermal state \cite{Son:2002sd,Herzog:2002pc}. An economical reformulation was developed by Glorioso, Crossley and Liu \cite{Glorioso:2018mmw}. One complexifies the radial coordinate and lets its contour encircle the horizon. The two sides of the radial contour represent the two Schwinger--Keldysh branches, while the analytic continuation around the horizon implements their thermal relation. No independent boundary condition is imposed at the horizon beyond requiring a single analytically continued bulk field on the full contour. The construction also organizes the derivative expansion
while keeping track of the horizon logarithms and their
monodromy.

We apply this construction to a minimally coupled massless probe scalar on a fixed Schwarzschild background in the Hartle--Hawking state. We express the finite-frequency problem in terms of exact ingoing and outgoing radial solutions, for which a Mano--Suzuki--Takasugi (MST) representation \cite{Mano:1996vt}  is available, and use a separate low-frequency expansion for tidal matching. Our first result is a direct contour derivation of \eqref{eq:fdtintro} for the finite-radius Dirichlet kernels, valid for every scalar multipole and arbitrary real frequency within the quadratic field problem. The relation itself is guaranteed by the KMS property of the Hartle--Hawking state~\cite{Higuchi:2021sxn}, and fluctuation--dissipation relations for black holes go back to Ref.~\cite{Candelas:1977zz}; here it is obtained directly from the radial construction. At leading order in the low-frequency expansion, the tidal kernels inherit the same relation. The Bose--Einstein factor is generated by the monodromy of the radial solutions together with the imaginary Schwarzschild-time shift between the two branches. The KMS relation is therefore encoded in the analytic continuation defining the Hartle--Hawking contour.

Our second result concerns the running relation \eqref{eq:645}. We show directly in the low-frequency radial problem, for every scalar multipole, that absorption and logarithmic running involve the same horizon amplitude of the static solution, while their ratio is fixed by the horizon slope $f'(\rs)=2\kappa$, with $\kappa$ the surface gravity. The factor $\rs=1/(2\kappa)$ in \eqref{eq:645} therefore enters through the radial geometry rather than through thermal occupation. This relation, derived in Sec.~\ref{sec:running}, agrees with the generic-multipole result of Ref.~\cite{Combaluzier:2025eoc}. Since it involves
only the retarded kernel, which is state-independent in this scalar field model, it should also hold in the Boulware vacuum, where there is no thermal occupation. It relates coefficients within the retarded response, whereas the thermal fluctuation--dissipation theorem \eqref{eq:fdtintro} additionally involves the state-dependent symmetric noise correlator.

The fluctuation--dissipation theorem and the running relation are thus closely connected, but they are not identical. Their low-frequency connection involves the same leading absorptive horizon datum and the natural dimensionless frequency
\begin{equation}
\eta=i\omega\rs=\frac{i\beta\omega}{4\pi},
\label{eq:etaIntro}
\end{equation}
where $\beta=4\pi\rs$ is the inverse Hawking temperature. At leading low frequency, KMS analyticity relates that absorptive property to thermal noise, while the near-zone radial evolution converts it into logarithmic conservative running. This common input explains the fluctuation--dissipation-type structure observed in \cite{Combaluzier:2025eoc,Rodriguez:2026iot}, but without directly identifying the renormalization group relation with the thermal fluctuation--dissipation theorem.

The paper is organized as follows. In Sec.~\ref{sec:schw} we define the response and noise kernels, formulate the scalar problem on the complex Schwarzschild radial contour, and determine its horizon monodromy and its exact and static radial solutions. Section~\ref{sec:fdtexact} derives the exact fluctuation--dissipation theorem. In Sec.~\ref{sec:running} we perform the tidal matching, derive the relation between leading absorption and the logarithmic running of the dynamical Love coefficient, and clarify its connection with the fluctuation--dissipation theorem. We conclude in
Sec.~\ref{sec:disc}. Appendix~\ref{app:kms} verifies dynamical KMS invariance of the quadratic functional.
Appendix~\ref{sub:mst} compares the leading running relation with the resummed response and the scalar
Raman-scattering matching. Appendix~\ref{sub:sourceNormalization} discusses source normalization and its connection to low-frequency absorption cross sections.

\section{The Schwarzschild scalar and the thermal contour}
\label{sec:schw}

\subsection{Sources, response, and noise}
\label{sec:inin}
\noindent
We first review and set up the general in--in description of the response and noise kernels and then specialize to the Schwarzschild scalar. We want to determine the expectation value, in a specified quantum state, of the multipole moment induced on a black hole by an applied external field. For definiteness, the external field will be a massless scalar. Throughout the main text we consider an eternal Schwarzschild black hole in the Hartle--Hawking thermal state~\cite{Hartle:1976tp}, working in the probe-field approximation and neglecting the backreaction of the thermal bath. Two features of the problem make the in--in formalism particularly appropriate. The first is causality: linear response is governed by a retarded correlator, whereas the in--out formalism naturally produces time-ordered correlators. The second is that the exterior region is an open system. Tracing over the inaccessible degrees of freedom gives reduced dynamics that need not be unitary. In thermal equilibrium, these dynamics exhibit both dissipation and fluctuations, which should be described within the same real-time framework.

\begin{figure}
\centering
\includegraphics[width=0.75\linewidth]{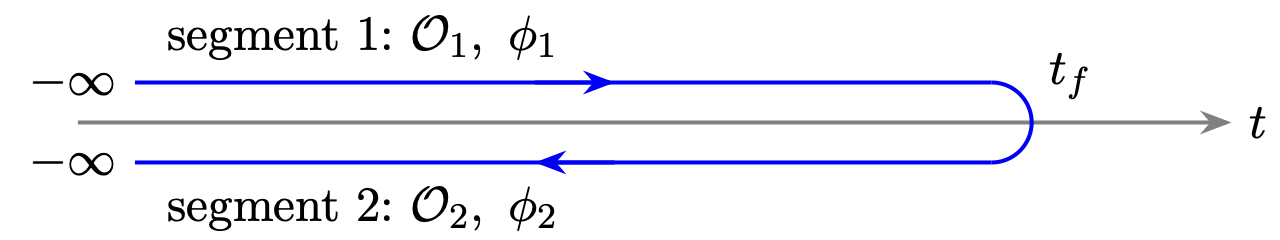}
\caption{The Schwinger--Keldysh closed-time-path contour.  Operators inserted on the upper and lower segments are labelled   by 1 and 2, respectively. The minus sign on branch 2 in \eqref{eq:Wgen} follows from the reversed orientation of the backward time evolution. The initial density matrix specifies the quantum state.}
\label{fig:sk}
\end{figure}

The appropriate framework is the closed-time-path, or in--in, formalism of Schwinger and Keldysh~\cite{Schwinger:1960qe,Keldysh:1964ud} (see \cite{Chou:1984es,Kamenev:2011ba} for reviews). One evolves forward from an initial time to a final time later than all operator insertions, and then backward to the initial time, as in Fig.~\ref{fig:sk}. The source is thereby doubled, with $\ph_1$ on the forward branch and $\ph_2$ on the backward branch. The generating functional is
\begin{align}
e^{iW[\ph_1,\ph_2]}=\operatorname{Tr}\!\left[U_{\ph_1}(t_f,t_0)\,\rho_0\,U_{\ph_2}^{\dagger}(t_f,t_0)\right]= \operatorname{Tr}\!\left[\rho_0\,\mathcal P e^{i\int_{t_0}^{t_f}\dd t\,(\mathcal O_1\ph_1-\mathcal O_2\ph_2)} \right].
\label{eq:Wgen}
\end{align}
Here
\begin{equation}
U_{\ph_j}(t_f,t_0)=\mathcal T\exp\!\left[i\int_{t_0}^{t_f}\dd t\,\ph_j(t)\mathcal O(t)\right]
\end{equation}
is the interaction-picture evolution operator for $H_{{\rm int},j}(t)=-\ph_j(t)\mathcal O(t)$. The normalized initial density matrix is $\rho_0$, and $\mathcal P$ denotes ordering along the closed time path. Here $\mathcal O$ is the multipole operator and $\ph$ is its conjugate external source; multipole labels and their contraction are temporarily suppressed. The operators
$\mathcal O_1$ and $\mathcal O_2$ denote insertions of the same physical operator on the two branches. The minus sign on branch 2 reflects the reversed orientation of the backward time evolution. The time variable $t$ is Schwarzschild time, which is also used in the stationary bulk calculation of Sec.~\ref{sub:scalar}. In the free bulk model used below, the normalized generating functional $W$ is obtained from the renormalized on-shell action, with the convention $e^{iW}$.

Setting the two sources equal makes the forward and backward evolutions cancel. For $\operatorname{Tr}\rho_0=1$, the generating functional then satisfies $W[\ph_1=\ph_2]=0$. This is the Schwinger--Keldysh normalization condition following from unitarity of the full theory   \cite{Crossley:2015evo,Liu:2018kfw}. It motivates the Keldysh combinations
\begin{equation}
\ph_r:=\tfrac12(\ph_1+\ph_2), \quad \ph_a:=\ph_1-\ph_2 ,
\quad
\phi_1\mathcal O_1-\phi_2\mathcal O_2=\phi_r(\mathcal O_1-\mathcal O_2)+\frac{\phi_a}{2}(\mathcal O_1+\mathcal O_2).
\label{eq:rabasis}
\end{equation}
The condition $W[\ph_r,\ph_a=0]=0$ implies that every term in the source expansion of $W$ carries at least one power of $\ph_a$. In particular, it excludes an $\ph_r\ph_r$ term. At quadratic order, the remaining structures are parametrized by a causal response kernel $K$ and a symmetric noise kernel $F$.

For a spherically symmetric background, we decompose the source into real orthonormal spherical harmonics,
\begin{equation}\label{spherical}
\ph(t,\Omega)=\sum_{\ell m}\ph_{\ell m}(t)Y_{\ell m}(\Omega),\quad
\phi_{\ell m}(t)=\int d\Omega\,Y_{\ell m}(\Omega)\phi(t,\Omega),
\quad \int\dd\Omega\, Y_{\ell m}Y_{\ell' m'}=\delta_{\ell\ell'}\delta_{mm'} .
\end{equation}
Using a real basis makes each $\ph_{\ell m}(t)$ real and avoids implicit complex-harmonic contractions.   Rotational invariance of the background and the Hartle--Hawking state makes the kernels diagonal in these labels and independent of $m$; i.e. $K_{\ell m,\ell'm'}(t,t')
=\delta_{\ell\ell'}\delta_{mm'}K_\ell(t,t')$ and  $F_{\ell m,\ell'm'}(t,t')
=\delta_{\ell\ell'}\delta_{mm'}F_\ell(t,t')$. Time-translation invariance additionally implies
$K_\ell(t,t')=K_\ell(t-t')$, and  $F_\ell(t,t')=F_\ell(t-t')$.  With the Fourier convention
\begin{equation}
\ph(\omega)=\int\dd t\,e^{i\omega t}\ph(t), \qquad \ph(t)=\int\frac{\dd\omega}{2\pi}\,e^{-i\omega t}\ph(\omega),
\label{eq:FourierConvention}
\end{equation}
we write
\begin{equation}
W=\sum_{\ell m}\int\!\frac{\dd\omega}{2\pi} \left[\ph_{a,\ell m}(-\omega)\,K_\ell(\omega)\,\ph_{r,\ell m}(\omega) +\frac{i}{2}\,F_\ell(\omega)\,\ph_{a,\ell m}(-\omega)\ph_{a,\ell m}(\omega)\right]+O(\ph^3).
\label{eq:Wdef}
\end{equation}
The kernels $K_\ell$ and $F_\ell$ characterize, respectively, the linear response and equilibrium fluctuations. We suppress the $\ell m$ labels on the sources and operators below whenever we
consider a single mode.

The response kernel is obtained by differentiating the generating functional. The two sources are identified after differentiation,
\begin{equation}
\langle\mathcal O(t)\rangle=\left.\frac{\delta W}{\delta\ph_a(t)}\right|_{\ph_a=0},
\qquad
\langle\mathcal O(\omega)\rangle=K_\ell(\omega)\,\ph_r(\omega)+{\cal O}(\ph_r^2),
\label{eq:Kdefn}
\end{equation}
where the expectation value in the absence of a source vanishes for the scalar response considered here.

The source coupling in \eqref{eq:Wgen} corresponds to $H_{\rm int}=-\mathcal O\ph$. Linear response therefore gives
\begin{equation}
K_\ell(\omega)=+i\int\dd t\,e^{i\omega t}\,\theta(t)\, \langle[\mathcal O(t),\mathcal O(0)]\rangle,
\label{eq:Kret}
\end{equation}
with the expectation value evaluated in the unperturbed thermal state. This sign is fixed by the source coupling and by \eqref{eq:Kdefn}. If the retarded Green function is instead defined as $G_R(t)=-i\theta(t)\langle[\mathcal O(t),\mathcal O(0)]\rangle$, then our response kernel is $K_\ell=-G_{R,\ell}$. It is $K_\ell$, with the convention \eqref{eq:Kret}, that appears in \eqref{eq:Wdef} and is matched to the tidal response coefficients in the worldline effective theory \cite{Goldberger:2005cd,Goldberger:2020fot,Goldberger:2020wbx,Rodriguez:2026iot}.

For a Hermitian multipole operator, $K_\ell(-\omega)=K_\ell(\omega)^*$. Thus ${\rm Re}\,K_\ell$ is even in $\omega$, whereas ${\rm Im}\,K_\ell$ is odd. The real part describes the conservative response, from which static and dynamical Love coefficients are extracted after tidal matching, while the imaginary part describes dissipation. With the conventions of \eqref{eq:Kret}, positive-frequency absorption corresponds to
\begin{equation}
{\rm Im}\,K_\ell(\omega)>0, \qquad \omega>0.
\label{eq:absign}
\end{equation}
For example, a monochromatic source $\ph_r(t)={\rm Re}(\ph_\omega e^{-i\omega t})$
delivers the time-averaged power
$\overline{\mathcal P} =\frac{\omega}{2}{\rm Im}\,K_\ell(\omega)|\ph_\omega|^2$, 
which is positive for an absorptive response with this convention.

The second kernel in \eqref{eq:Wdef} is the connected symmetric correlator,
\begin{equation}
F_\ell(t)=\frac12 \left\langle \left\{\delta\mathcal O(t),\delta\mathcal O(0)\right\} \right\rangle,
\qquad
\delta\mathcal O\equiv \mathcal O-\langle\mathcal O\rangle .
\label{eq:Fdef}
\end{equation}
Its Fourier transform uses \eqref{eq:FourierConvention}.
In the unperturbed state considered here, $\delta\mathcal O=\mathcal O$.
The kernel $F_\ell(\omega)$ is real, even in frequency, and non-negative. In $e^{iW}$, the imaginary quadratic term in \eqref{eq:Wdef} therefore produces the Gaussian damping factor
\begin{equation}
\exp\left[ -\frac12\int\frac{\dd\omega}{2\pi}\, F_\ell(\omega)\, \ph_a(-\omega)\ph_a(\omega) \right].
\end{equation}
A Hubbard--Stratonovich transformation represents this factor through a real Gaussian stochastic force $\xi$ satisfying
$\langle\xi(\omega)\xi(\omega')\rangle_\xi = 2\pi\delta(\omega+\omega')F_\ell(\omega)$
\cite{Feynman:1963fq,Kamenev:2011ba}.
This stochastic representation reproduces the symmetric fluctuations at quadratic order. For the black-hole response, these include the equilibrium quantum and thermal fluctuations of the degrees
of freedom integrated out in the effective description.

In thermal equilibrium, the KMS condition relates the symmetric and response correlators through the fluctuation--dissipation theorem \eqref{eq:fdtintro}
\cite{Kubo:1957mj,Martin:1959jp} (see also Ref.~\cite{Mottola:1985ee}).  The positive sign follows from \eqref{eq:Kret} and, together with \eqref{eq:absign},  ensures positive noise; the expression at $\omega=0$ is understood as a limit. The thermal state is an input to this relation. 

In summary, in this subsection we formulated the response problem in the interaction picture using the Schwinger--Keldysh contour, with independent source histories on the forward and backward branches. At quadratic order in the sources, the connected generating functional $W$ is determined by two kernels: $K$, which describes linear response, including conservative effects and absorption, and $F$, which describes equilibrium fluctuations. Interaction-picture evolution and functional differentiation identify $K$ with the retarded commutator of the multipole operator $\mathcal O$, and $F$ with its connected symmetric anticommutator. Below we show how the corresponding KMS structure is implemented by the gravitational contour.

\subsection{Scalar field and boundary action}
\label{sub:scalar}
\begin{figure}
    \centering
    \includegraphics[width=0.7\linewidth]{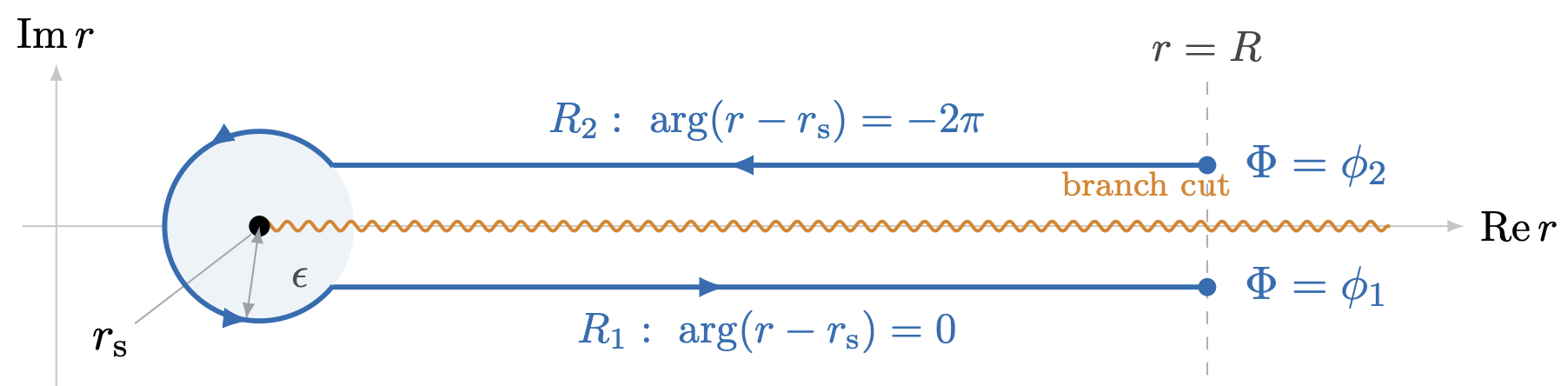}
    \caption{The complex radial contour joining the two
    Schwinger--Keldysh segments. The straight portions carry the
    boundary data at the two copies of the matching surface,
    while a circle of radius $\epsilon$ connects them around
    the horizon branch point. The continuation includes the
    corresponding imaginary-time identification described in
    Sec.~\ref{sub:horizon-continuation}.}
    \label{fig:contour}
\end{figure}

\noindent
We now specialize the Schwinger--Keldysh construction to a minimally coupled massless scalar on a fixed Schwarzschild background.
We write the background metric in Schwarzschild coordinates,
\begin{equation}
\dd s^2=-f(r)\,\dd t^2+\frac{\dd r^2}{f(r)}+\lambda_{ij}(r)\,\dd x^i\dd x^j,
\qquad
\lambda_{ij}=r^2\gamma_{ij}(\Omega).
\label{eq:metricS}
\end{equation}
Here $\gamma_{ij}$ is the unit-sphere metric. We absorb its volume element into $\dd\Omega$ and use $\sqrt{\lambda}=r^2$ for the
remaining radial area factor. For Schwarzschild,
\begin{equation}
f(r)=1-\frac{\rs}{r},\qquad \kappa=\frac12 f'(\rs)=\frac{1}{2\rs},
\qquad \beta=\frac{2\pi}{\kappa}=4\pi\rs, \qquad \AH=4\pi\rs^2,
\label{eq:schw}
\end{equation}
with $\rs=2GM$, $\Thaw=\beta^{-1}$, and $\sqrt{\lambda_h}=\rs^2$. The tortoise coordinate is
\begin{equation}
r_*=r+\rs\log\!\left(\frac{r}{\rs}-1\right), 
\label{eq:tortoise}
\end{equation}
which gives the convenient formula $\partial_r r_*=f^{-1}$. Following Ref. \cite{Glorioso:2018mmw}, we analytically continue the geometry around $r=\rs$ along the contour of Fig.~\ref{fig:contour}. The first and second sheets represent the two Schwinger--Keldysh branches. We orient the radial integration contour $\mathcal C$ from $R_2$ inward toward the horizon and then
outward to $R_1$. The inverse continuation, from sheet 1 to sheet 2, is clockwise: $\arg(r-\rs)\to\arg(r-\rs)-2\pi$.
The contour therefore winds counterclockwise in its integration orientation. At the two copies $R_1$ and $R_2$ of a finite radial surface $r=R$, we prescribe Dirichlet boundary sources $\varphi_1$ and $\varphi_2$.

The continuation is performed at fixed regular ingoing coordinate $v=t+r_*$. Consequently, Schwarzschild time is itself shifted between the two sheets. The Schwarzschild-coordinate expressions below are understood as expressions on this complexified geometry, with the induced time identification included. This distinction is essential when relating the Fourier amplitudes on the two
branches.

Using the real orthonormal harmonics introduced in \eqref{spherical},
\begin{equation}
\Phi(t,r,\Omega) = \sum_{\ell m}\Phi_{\ell m}(t,r)Y_{\ell m}(\Omega),\qquad\int\dd\Omega\, Y_{\ell m}Y_{\ell' m'}=\delta_{\ell\ell'}\delta_{mm'},
\label{eq:Ynorm}
\end{equation}
with $\nabla_{S^2}^2Y_{\ell m}=-\ell(\ell+1)Y_{\ell m}$,
the action for a fixed $(\ell,m)$ mode is
\begin{equation}
S=\frac12\int\dd t\int_{\mathcal C}\dd r \left[\frac{r^2}{f}(\partial_t\Phi)^2-f r^2 (\partial_r\Phi)^2-\ell(\ell+1)\Phi^2\right],
\label{eq:actionS}
\end{equation}
where mode labels are suppressed, $\partial_r$ is taken at fixed Schwarzschild time, and $\Pi:=-fr^2\partial_r\Phi$ is the radial canonical momentum. The time integration follows the corresponding analytically continued time contour.

Varying the Schwarzschild-coordinate action
\eqref{eq:actionS} gives
\begin{equation}
\mathcal L\Phi=\frac{r^2}{f}\,\partial_t^2\Phi,
\qquad \mathcal L\Phi \equiv \partial_r\!\left(f r^2\,\partial_r\Phi \right) -\ell(\ell+1)\Phi ,
\label{eq:eomS}
\end{equation}
which in the Fourier convention \eqref{eq:FourierConvention} becomes,
\begin{equation}
\mathcal L\Phi_\omega=-\frac{r^2}{f}\,\omega^2\Phi_\omega .
\label{eq:eomSomega}
\end{equation}
The radial equation depends on frequency only through $\omega^2$. Its odd-frequency response therefore arises from the causal
boundary condition rather than from an odd-frequency term in the Schwarzschild radial operator.

To evaluate the boundary terms, we return temporarily to ingoing coordinates, in which the radial contour is traversed at fixed $v$. This allows integration by parts without introducing the $r$-dependent Schwarzschild-time integration limits.  Introducing $v=t+r_*(r)$ and defining
$\widehat\Phi(v,r)\equiv\Phi(v-r_*(r),r)$, the same action \eqref{eq:actionS} is
\begin{equation}
S=-\frac12\int\dd v\int_{\mathcal C}\dd r \left[f r^2(\partial_r\widehat\Phi)^2+2r^2\partial_v\widehat\Phi\,\partial_r\widehat\Phi+\ell(\ell+1)\widehat\Phi^2\right],
\end{equation}
where $\partial_r\widehat\Phi$ is taken at fixed $v$. In Schwarzschild coordinates this same contour has $t=v-r_*(r)$. The coordinate transformation leaves its radial orientation unchanged. Using the equation of motion \eqref{eq:eomS}, expressed in these coordinates, the on-shell action reduces to
\begin{align}
S_{\rm os}={}-\frac12\int_{v_i}^{v_f}\dd v\,\left[\widehat\Phi
\left(r^2f\,\partial_r\widehat\Phi+r^2\partial_v\widehat\Phi\right)\right]_{R_2}^{R_1}
-\frac12\int_{\mathcal C}\dd r\,\left[r^2\widehat\Phi\,\partial_r\widehat\Phi\right]_{v_i}^{v_f},
\label{parts1}
\end{align}
where radial derivatives are taken at fixed $v$. The temporal endpoint terms cancel when the two Lorentzian branches and the thermal preparation contribution are combined with the Schwinger--Keldysh gluing conditions, leaving only the radial boundary term.

We now express the radial boundary term in Schwarzschild coordinates to use the same time coordinate and Fourier convention as in the radial equation \eqref{eq:eomSomega}.  At each fixed-radius endpoint, $\dd v=\dd t$, and the chain rule turns the combination in the first term of \eqref{parts1} into
\begin{equation}\label{chain01}
r^2 f\,\partial_r\widehat\Phi+r^2\partial_v\widehat\Phi=r^2 f \,\left.\partial_r\Phi\right|_t.
\end{equation}
This identity puts the boundary action in a form directly suited to extracting
the frequency-space response and noise kernels.

The connection to the generating functional follows from the normalized Schwinger--Keldysh path integral $Z=e^{iW}$ on the contour $\mathcal C$, with the
Hartle--Hawking state preparation and contour gluing. Integrating out the
bulk scalar with its boundary values held fixed,
\begin{equation}
\mathcal N\, Z[\varphi_1,\varphi_2]=\int_{\Phi|_{R_i}=\varphi_i}\mathcal D\Phi\;
e^{\,i\left(S_{\mathcal C}[\Phi]+S_{\rm ct}[\varphi_1,\varphi_2]\right)},
\end{equation}
where $S_{\rm ct}$ is the boundary counterterm and $\mathcal N$ fixes $Z[0,0]=1$.
Writing $\Phi=\Phi_{\rm cl}+\chi$, with $\Phi_{\rm cl}$ the classical solution with
Dirichlet data $\varphi_i$ and $\chi|_{R_i}=0$, the cross term vanishes on shell and
the quadratic action splits as $S_{\mathcal C}[\Phi]=S_{\rm os}[\varphi_1,\varphi_2]+S_{\mathcal C}[\chi]$.
The $\chi$ integral is source-independent and is removed by $\mathcal N$, so
\begin{equation}
W[\varphi_1,\varphi_2]=S_{\rm os}[\varphi_1,\varphi_2]+S_{\rm ct}[\varphi_1,\varphi_2].
\end{equation}
Using \eqref{chain01}, the on-shell action is the radial boundary term in \eqref{parts1},
\begin{equation}
W_{\rm bulk}\equiv S_{\rm os}=-\frac12\int\dd t\,\Big[f r^2\,\Phi\,\partial_r\Phi \Big]_{R_2}^{R_1}
=-\frac12\int\!\frac{\dd\omega}{2\pi}\,\Big[f\sqrt{\lambda}\,\Phi_{-\omega}\,\partial_r\Phi_\omega\Big]_{R_2}^{R_1}.
\label{eq:WbdyS}
\end{equation}
Writing $W_{\rm ct}\equiv S_{\rm ct}$ for the local counterterms required by tidal matching, $W=W_{\rm bulk}+W_{\rm ct}$. The on-shell variation of the renormalized action is
\begin{equation}
\delta(S_{\rm os}+S_{\rm ct})
=\int\dd t\,\left[\Pi_{1,\rm ren}\,\delta\varphi_1-\Pi_{2,\rm ren}\,\delta\varphi_2\right],
\end{equation}
where the renormalized endpoint momenta $\Pi_{j,\rm ren}$ include the boundary-counterterm contribution. This has the same source-variation structure as the Schwinger--Keldysh generating functional \eqref{eq:Wgen}: the renormalized endpoint momenta provide the bulk realization of the operators conjugate to the boundary sources.

In summary, we have constructed a bulk realization of the generating functional $W$ introduced in the previous subsection, with boundary field values as sources and renormalized boundary momenta giving the expectation values of their conjugate operators. The remaining task is to solve the bulk boundary-value problem
and determine the endpoint momenta $\Pi$ as functionals of the prescribed boundary sources $\varphi_1$ and $\varphi_2$.

\subsection{Horizon behaviour and thermal continuation}
\label{sub:horizon-continuation}
\noindent
Let us introduce the dimensionless coordinate
\begin{equation}
y=\frac{r-\rs}{\rs},\qquad f(r)=\frac{y}{1+y}.
\end{equation}
In this coordinate, the exact radial equation
\eqref{eq:eomSomega} becomes
\begin{equation}
\partial_y\!\left[y(1+y)\partial_y\Phi_\omega\right]+\left[
(\omega\rs)^2\frac{(1+y)^3}{y}-\ell(\ell+1)\right]\Phi_\omega=0.
\label{eq:heun}
\end{equation}
Near the horizon, $y\to0$ and $f\simeq y$.
The indicial equation is
$\alpha^2+(\omega\rs)^2=0$, so the two Frobenius
solutions behave as
\begin{equation}
\Phi_\omega\simeq \mathcal A_+y^{i\omega \rs} +\mathcal A_-y^{-i\omega \rs} .
\label{eq:varpi}
\end{equation}
The powers have dimensionless arguments. From \eqref{eq:tortoise}: $y^{\pm i\omega \rs}=e^{\pm i\omega r_*}e^{\mp i\omega r}.$
The final factor is regular and nonzero at the horizon. With the time dependence $e^{-i\omega t}$, the minus sign in \eqref{eq:varpi}, $y^{-i\omega \rs}$ is therefore ingoing and the plus sign $y^{i\omega \rs}$ outgoing. The centrifugal term is subleading in the indicial equation, so these exponents are independent of $\ell$. The retarded problem selects the ingoing solution, with no independently supplied outgoing signal from the horizon.   Writing $y=|y|e^{i\theta}$, clockwise continuation from sheet 1 to sheet 2 gives $\theta\to\theta-2\pi$, hence $\log y\to\log y-2\pi i$, and the radial horizon factors transform as
\begin{equation}
y^{+i\omega \rs} \to e^{+\beta\omega/2} y^{+i\omega \rs},
\qquad
y^{-i\omega \rs} \to e^{-\beta\omega/2}y^{-i\omega \rs}.
\label{eq:monoS}
\end{equation}

The same clockwise continuation shifts the tortoise coordinate \eqref{eq:tortoise}:
$r_*=r+\rs\log y,$ by $r_*\to r_*-2\pi i\rs=r_*-i\beta/2$. At fixed $v=t+r_*$, this implies
\begin{equation}
t_2=t_1+\frac{i\beta}{2}.
\label{eq:tKMS}
\end{equation}
The time factor $e^{-i\omega t}$ thus acquires $e^{+\beta\omega/2}$. Combined with \eqref{eq:monoS}, this leaves the full ingoing mode unchanged and multiplies the full outgoing mode by $e^{\beta\omega}$. These are the continuation factors used in the Schwinger--Keldysh boundary-value problem below. The thermal interpretation uses both this analytic continuation and the Hartle--Hawking state prescription and it does not follow from the radial exponents alone. 

\subsection{Exact and static radial solutions}
\label{sec:exact}
\noindent
Let $\psi_\pm(r,\omega)$ denote the exact exterior solutions of \eqref{eq:eomSomega} that are respectively outgoing and
ingoing at the horizon, normalized for nonzero real frequency by
\begin{equation}
\psi_\pm(r,\omega)=\left(\frac{r-\rs}{\rs}\right)^{\pm i\omega \rs} \left[1+O\!\left(\frac{r-\rs}{\rs}\right)\right].
\label{eq:psipm}
\end{equation}
At fixed exterior radius, both normalized solutions approach $P_\ell(2r/\rs-1)$ as $\omega\to0$.
The second independent static solution is recovered from a rescaled difference, as discussed below. The radial equation \eqref{eq:heun} is of confluent-Heun type. The MST construction \cite{Mano:1996vt} provides an exterior representation in bilateral series of Coulomb wave functions \cite{Kosmopoulos:2025rfj}. In our analysis we use only the existence of $\psi_\pm$ for real $\omega$ and their horizon normalization, not the explicit series.

For real $\omega$, the radial equation has real coefficients on the physical exterior and depends
on frequency only through $\omega^2$. Together with the unit horizon normalization \eqref{eq:psipm}, these properties imply
\begin{equation}
\psi_-(r,\omega)=\psi_+(r,-\omega),\qquad \psi_+(r,\omega)=\psi_-(r,\omega)^*.
\label{eq:crossing}
\end{equation}
A real field on the physical Lorentzian exterior satisfies $\Phi_{-\omega}(r)=\Phi_\omega(r)^*$. The finite-radius absorption will be fixed by the conserved radial flux of the scalar field, which we now introduce. For two functions of $r$ we define the weighted Wronskian
\begin{equation}
{\mathcal W}[\psi_1,\psi_2] := f r^2 \left(\psi_1\partial_r\psi_2- \psi_2\partial_r\psi_1 \right) . 
\label{eq:Wdeff1}
\end{equation}
For two solutions with the same frequency and multipole, differentiating and using the radial equation \eqref{eq:eomSomega}, we obtain $\partial_r\mathcal W=\psi_1\partial_r(fr^2\partial_r\psi_2)-\psi_2\partial_r(fr^2\partial_r\psi_1)=0,$ 
so $\mathcal W$ is independent of $r$. Evaluating it at the horizon using \eqref{eq:psipm} gives for the conserved Wronskian \eqref{eq:Wdeff1}
\begin{equation}
\mathcal W[\psi_-,\psi_+]=2i\omega\rs^2.
\label{eq:Whor}
\end{equation}
Thus, radial conservation fixes the Wronskian at every exterior radius to this exact horizon value when the horizon limit is taken at fixed $\omega$. 

For the later tidal matching, we also need a static
basis that distinguishes the applied growing field
from the induced decaying field. Setting $\omega=0$
and introducing $x=2r/\rs-1$, the radial equation   \eqref{eq:eomSomega}  becomes
\begin{equation}\label{eq:eomlegendre}
\partial_x\!\left[(x^2-1)\partial_x\Phi\right]
-\ell(\ell+1)\Phi=0.
\end{equation}
Its independent solutions are $P_\ell(x)$ and $Q_\ell(x)$, with $Q_\ell$ chosen real for $x>1$ on the physical sheet.
At large $x$, $P_\ell(x)\sim p_\ell x^\ell$ and $Q_\ell(x)\sim q_\ell x^{-\ell-1}$, where
\begin{equation}
p_\ell=\frac{(2\ell)!}{2^\ell(\ell!)^2},
\qquad
q_\ell=\frac{2^\ell(\ell!)^2}{(2\ell+1)!},
\qquad
p_\ell q_\ell=\frac{1}{2\ell+1}.
\label{eq:legasym}
\end{equation}
Their Wronskian is
\begin{equation}
(x^2-1)\left(P_\ell Q_\ell'-P_\ell'Q_\ell\right)=-1,
\label{eq:legWronskian}
\end{equation}
where primes in \eqref{eq:legWronskian} denote $\partial_x$.
Thus,
\begin{equation}
\mathcal W[P_\ell,Q_\ell]=-\frac{\rs}{2}.
\label{eq:WPQ1}
\end{equation}

Having identified the growing and decaying modes at large radius, we now examine their behaviour at the horizon, $x=1$.
The solution $P_\ell$ is regular there, with $P_\ell(1)=1$, whereas the independent solution $Q_\ell$ has a logarithmic
branch point:
\begin{equation}
Q_\ell(x)=\frac12P_\ell(x)\log\frac{x+1}{x-1}-w_{\ell-1}(x),
\label{eq:Qlog}
\end{equation}
where $w_{\ell-1}$ is a polynomial  of degree $\ell-1$ and is absent for $\ell=0$.  Clockwise continuation around $x=1$, without
encircling $x=-1$, gives
\begin{equation}
Q_\ell\longrightarrow Q_\ell+i\pi P_\ell, \qquad P_\ell\longrightarrow P_\ell .
\label{eq:legmono}
\end{equation}
The logarithmic static solution arises when the two horizon exponents $\pm i\omega r_s$ coalesce at zero frequency:
the rescaled difference of the corresponding solutions retains a $\log[(r-r_s)/r_s]$ term. It therefore reproduces $Q_\ell$,
up to normalization and addition of the regular solution $P_\ell$. Its additive monodromy is the zero-frequency counterpart of the
multiplicative continuation factors in \eqref{eq:monoS}.

In this subsection, we have defined the exact radial solutions and their conserved Wronskian, then obtained the static solutions $P_\ell$ and $Q_\ell$ by setting $\omega=0$. Their large-radius behaviour identifies the growing and decaying modes needed for tidal matching. We have also examined their horizon continuation: $P_\ell$ remains unchanged, while $Q_\ell$ acquires $i\pi P_\ell$, establishing the additive static monodromy.

\section{The fluctuation--dissipation theorem}
\label{sec:fdtexact}
\noindent
We now return to the exact finite-frequency solutions and derive the fluctuation--dissipation theorem at a finite radial surface. 
The thermal input is the Hartle--Hawking contour specified in the previous Sec.~\ref{sec:schw}. Tidal matching will be performed only after obtaining the exact boundary kernels.

Let $\psi_\pm$ be the outgoing and ingoing solutions normalized by \eqref{eq:psipm}. On sheet 1, define
\begin{equation}
g_\pm(r,\omega):=\frac{\psi_\pm(r,\omega)}{\psi_\pm(R_1,\omega)},\qquad g_\pm(R_1,\omega)=1,
\label{eq:gpm}
\end{equation}
and write
\begin{equation}
\Phi_\omega(r)=X(\omega)g_+(r,\omega)+Y(\omega)g_-(r,\omega).
\label{eq:PhiXY}
\end{equation}
Here $\Phi_\omega(r)$ is the Schwarzschild-time radial mode function amplitude on sheet 1, continued analytically to sheet 2. The normalization denominators in \eqref{eq:gpm} are held fixed at their sheet 1 values during continuation.

Using $v=t+r_*(r)$, define the common real boundary-time parameter $\tau=v-r_*(R_1)$, which equals Schwarzschild time on
sheet~1 at the boundary. Then
\begin{equation}
e^{-i\omega t}\Phi_\omega(r)=e^{-i\omega\tau}e^{i\omega[r_*(r)-r_*(R_1)]}\Phi_\omega(r).
\end{equation}
At $R_1$, the radial phase factor
$e^{i\omega[r_*(r)-r_*(R_1)]}$ equals unity. At $R_2$, \eqref{eq:tKMS} gives $t_2=\tau+i\beta/2$. The finite-radius Dirichlet sources are therefore $\varphi_1(\tau)=\Phi(\tau,R_1),$
$\varphi_2(\tau)=\Phi(\tau+i\beta/2,R_2),$
where the second expression uses the analytically continued bulk field. The sources $\varphi_j(\tau)$ are the ordinary real-time Schwinger--Keldysh sources. Their Fourier transforms with respect to $\tau$ satisfy 
\begin{equation}\label{eq:phi1phi2}
\varphi_1(\omega)=\Phi_\omega(R_1), \qquad \varphi_2(\omega)=e^{\beta\omega/2}\Phi_\omega(R_2).
\end{equation}
The momentum amplitudes acquire the same time-conversion factor. These factors cancel
between the opposite frequencies in each boundary bilinear, leaving \eqref{eq:WbdyS} unchanged.

The radial continuation \eqref{eq:monoS} gives
$g_\pm(R_2,\omega)=e^{\pm\beta\omega/2}$. Combining these factors with the time identification, we obtain
\begin{equation}
\varphi_1=X+Y, \qquad \varphi_2=e^{\beta\omega}X+Y.
\label{eq:cpmSources}
\end{equation}
For presentation reasons we suppress the frequency arguments when unambiguous. 
Solving for the coefficients gives
\begin{equation}
X=-n(\omega)\varphi_a, \qquad Y=\varphi_r+ \frac12\coth\!\left(\frac{\beta\omega}{2}\right)\varphi_a, \qquad
n(\omega):=\frac{1}{e^{\beta\omega}-1},
\label{eq:bose}
\end{equation}
where  $\varphi_r,\varphi_a$ are defined as in \eqref{eq:rabasis}. 
Equal sources, $\varphi_a=0$, therefore select the purely ingoing solution $\Phi_\omega=\varphi_r g_-$. For unequal sources, the individual coefficients $X$
and $Y$ can diverge as $\omega\to0$. Nevertheless, $g_--g_+={\cal O}(\omega)$ at fixed exterior radius, so substituting \eqref{eq:bose} into \eqref{eq:PhiXY} we obtain
\begin{equation}
\Phi_\omega=\varphi_r g_- +\varphi_a\left[\frac12 g_-+n(\omega)(g_--g_+)\right]
\end{equation}
which has a finite zero-frequency limit. This limit contains the static logarithmic solution $Q_\ell$ given in \eqref{eq:Qlog}.

To evaluate the boundary action \eqref{eq:WbdyS}, we must relate the endpoint momenta to the prescribed boundary fields. The radial canonical momentum  is $\Pi_\omega(r)=-f(r)r^2\partial_r\Phi_\omega(r),$
where the radial derivative is taken at fixed Schwarzschild time before Fourier transformation.
Using the normalized solutions \eqref{eq:gpm}, we define
\begin{equation}
\mathcal K_\pm(R,\omega):=-f(R)R^2\partial_r g_\pm(R,\omega)
=-f(R)R^2\frac{\partial_r\psi_\pm(R,\omega)}{\psi_\pm(R,\omega)}.
\label{eq:KpmExact}
\end{equation}
Since $g_\pm(R,\omega)=1$, these kernels give the boundary momentum per unit Dirichlet amplitude for the outgoing and ingoing solutions, respectively. For real frequency, the crossing relations \eqref{eq:crossing} and the reality of $f(R)$ imply
\begin{equation}
\mathcal K_+(R,\omega)=\mathcal K_-(R,-\omega)=\mathcal K_-(R,\omega)^*.
\label{eq:KpmCross}
\end{equation}
Substituting \eqref{eq:PhiXY} into the momentum gives
\begin{equation}
\Pi_\omega(r)=-f(r)r^2\left[X\,\partial_r g_+(r,\omega)+Y\,\partial_r g_-(r,\omega)\right].
\end{equation}
At $R_1$, this yields $\Pi_1=\mathcal K_+X+\mathcal K_-Y$. Since the monodromy factors are independent of $r$, the radial derivatives acquire the same factors as $g_\pm$. Including the time shift \eqref{eq:tKMS}, as in \eqref{eq:cpmSources}, therefore gives
\begin{equation}
\Pi_1=\mathcal K_+X+\mathcal K_-Y, \qquad \Pi_2=e^{\beta\omega}\mathcal K_+X+\mathcal K_-Y.
\label{eq:Pi12}
\end{equation}
Here $\Pi_1$ and $\Pi_2$ are the endpoint momentum
amplitudes Fourier transformed with respect to the
common boundary-time convention introduced above.

The boundary action \eqref{eq:WbdyS} can then be expressed in terms of momenta as 
\begin{equation}
W=\frac12\int\!\frac{\dd\omega}{2\pi}\left[\varphi_1(-\omega)\Pi_1(\omega)-\varphi_2(-\omega)\Pi_2(\omega)\right]+W_{\rm ct}.
\label{eq:WexactPi}
\end{equation}
Substituting \eqref{eq:bose} and \eqref{eq:Pi12}
gives the endpoint bilinear
\begin{align}
\varphi_1(-\omega)\Pi_1(\omega)-\varphi_2(-\omega)\Pi_2(\omega)
={}&\varphi_a(-\omega)\mathcal K_-\varphi_r(\omega)+\varphi_r(-\omega)\mathcal K_+\varphi_a(\omega)
\nonumber\\
&+
\frac12\coth\!\left(\frac{\beta\omega}{2}\right)(\mathcal K_--\mathcal K_+)\varphi_a(-\omega)\varphi_a(\omega).
\end{align}
There is no $\varphi_r\varphi_r$ term, as required by Schwinger--Keldysh normalization.
Relabelling $\omega\to-\omega$ in the second term and using \eqref{eq:KpmCross}, we bring the action into the desirable form \eqref{eq:Wdef}
\begin{equation}
W=\int\!\frac{\dd\omega}{2\pi}\left[\varphi_a(-\omega)K_\ell^R(\omega)\varphi_r(\omega)+\frac{i}{2}F_\ell^R(\omega)
\varphi_a(-\omega)\varphi_a(\omega)\right],
\label{eq:WexactRA}
\end{equation}
where the superscript $R$ denotes the finite-radius
source normalization, and
\begin{equation}
K_\ell^R(\omega)=\mathcal K_-(R,\omega)+K_{\ell,\rm ct}(R,\omega),
\qquad
F_\ell^R(\omega)=\coth\!\left(\frac{\beta\omega}{2}\right){\rm Im}\,\mathcal K_-(R,\omega).
\label{eq:KExactCutoff}
\end{equation}
The conservative local counterterms used here contribute a real, even-in-frequency kernel $K_{\ell,\rm ct}$. Thus they do not change the absorptive part, so
\begin{equation}
F_\ell^R(\omega)=\coth\!\left(\frac{\beta\omega}{2}\right){\rm Im}\,K_\ell^R(\omega).
\label{eq:fdtexact}
\end{equation}
This relation holds for every multipole and real frequency within the stationary quadratic scalar problem, with $\omega=0$ understood by continuity. The thermal contour generates the Bose--Einstein factor $n(\omega)$ \eqref{eq:bose}, and hence the factor $1+2n(\omega)$ in \eqref{eq:fdtexact}, without an additional horizon matching condition, once the thermal state is chosen. 

To verify the absorptive sign and identify how propagation from the horizon affects the finite-radius response, we express
${\rm Im}\,\mathcal K_-$ in terms of the conserved radial Wronskian. Since $\mathcal K_+=\mathcal K_-^*$,
\eqref{eq:KpmExact} gives
\begin{equation}
\mathcal K_-(R,\omega)-\mathcal K_+(R,\omega)=\frac{\mathcal W[\psi_-,\psi_+]}
{|\psi_-(R,\omega)|^2}.
\end{equation}
Using \eqref{eq:Whor}, we obtain
\begin{equation}
{\rm Im}\,\mathcal K_-(R,\omega)=\frac{\omega\rs^2}{|\psi_-(R,\omega)|^2}.
\label{eq:ImKcutoff}
\end{equation}
This is positive for $\omega>0$, consistently with \eqref{eq:absign}. Propagation from the horizon to $R$ changes the magnitude of both absorption and noise, but cancels from their ratio in \eqref{eq:fdtexact}.

As a complementary check, Appendix~\ref{app:kms} shows that the quadratic functional is invariant under the dynamical KMS transformation precisely when its response and noise kernels satisfy \eqref{eq:fdtexact}.

In this section we have expressed the bulk solution in terms of the two Schwinger--Keldysh boundary sources, with the thermal continuation introducing the Bose--Einstein factor. We then wrote the boundary functional $W$ in terms of the endpoint momenta $\Pi_1,\Pi_2$, expressed through the kernels $\mathcal K_\pm$. Bringing $W$ into the form \eqref{eq:Wdef} allowed us to identify the response kernel $K$ and noise kernel $F$, and establish the equilibrium fluctuation--dissipation theorem at finite radius, exactly in frequency.

\section{Tidal matching and the running of the dynamical Love numbers}
\label{sec:running}
\noindent
The exact kernels obtained above describe the response to prescribed field values at a finite radius. To extract the dynamical Love coefficients, we must instead separate the applied tidal field from the field generated by the induced multipole. This requires a low-frequency matching calculation.

\subsection{Tidal matching}
\label{sub:tidal-matching}
\noindent
The Dirichlet source $\varphi_j(\omega)$ is the total scalar field at the matching surface $r=R$ on branch $j$, expressed in the common boundary-time convention introduced above. It therefore contains both the applied field and the induced response. The tidal source, by contrast, is the amplitude of the applied growing mode. We now relate these two source definitions by decomposing the low-frequency radial solution into its growing and decaying components.

Identifying an applied tide and an induced multipole requires a low-frequency expansion. For $|\omega|\rs\ll1$, we choose a matching surface in the overlap region,
\begin{equation}\label{eq:region}
\rs\ll R\ll\frac{1}{|\omega|}.
\end{equation}
Here the radius is large compared with the black hole scale, so the applied growing component and the induced decaying component can be identified through their respective $r^\ell$ and $r^{-\ell-1}$ behaviours. At the same time, $|\omega|R\ll1$ permits a near-zone expansion of the radial solution.

To relate the resulting tidal amplitudes to the Dirichlet sources used in the boundary functional, both must be expressed in the same boundary-time convention. We therefore multiply the Schwarzschild-time radial amplitude on each branch by
the corresponding $r$-independent time-conversion factor introduced in \eqref{eq:phi1phi2}. Since this factor is constant in $r$, the rescaled amplitude satisfies the same radial equation. 

The low-frequency expansion requires separate care near the horizon. Expanding the horizon factors $y^{\pm i\omega\rs}$ generates powers of $\omega\rs\log y$. Consequently, the condition $|\omega|\rs\ll1$ alone does not control the expansion arbitrarily close to the horizon. On the regulator circle, we have
\begin{equation}
r-\rs=\epsilon e^{i\theta}, \qquad \log y=\log\frac{\epsilon}{\rs}+i\theta.
\end{equation}
If the horizon factors are expanded on this circle, uniform control over a full turn requires
\begin{equation}
\frac{\epsilon}{\rs}\ll1,
\qquad
|\omega|\rs\left|\log\!\left(\frac{\epsilon}{\rs}\right)\right|\ll1.
\label{eq:epsiloncond}
\end{equation}
The first condition places the circle near the horizon, while the second controls the Taylor expansion of the horizon factors in frequency.  Together they imply $|\omega|\rs\ll1$, which also controls the finite angular variation of $\log y$ over a full turn. The second condition is unnecessary when the horizon factors are kept unexpanded. Any subsequent $\epsilon\to0$ limit is taken only after evaluating and combining the contour contributions at a fixed order in the frequency expansion, provided that the resulting coefficient has a finite limit. Thus the expansion and regulator limits are ordered; the truncated expansion is not uniform as $\epsilon\to0$ at fixed nonzero frequency. These restrictions concern the use of a frequency-expanded solution on the regulated Schwinger--Keldysh contour. Notice that the exact retarded kernel, obtained from the ingoing solution on sheet~1, does not depend on $\epsilon$.

Expanding the radial solution at fixed radius, $\Phi_\omega=\Phi^{[0]}+\omega\Phi^{[1]}+\omega^2\Phi^{[2]}+\cdots,$
and substituting into \eqref{eq:eomSomega} gives
\begin{equation}
\mathcal L\Phi^{[0]}=0, \qquad \mathcal L\Phi^{[1]}=0, \qquad \mathcal L\Phi^{[2]}=-\frac{r^2}{f}\Phi^{[0]}.
\end{equation}
Its independent homogeneous solutions are the static Legendre functions $P_\ell(x)$ and $Q_\ell(x)$ as in \eqref{eq:eomlegendre}. 
Thus
\begin{align}
\Phi^{[0]}&=A_0P_\ell+B_0Q_\ell,
&
\Phi^{[1]}&=A_1P_\ell+B_1Q_\ell,
\\
\Phi^{[2]}&=A_2P_\ell+B_2Q_\ell+\chi^{[2]},
&
\mathcal L\chi^{[2]}&=-\frac{r^2}{f}\Phi^{[0]}.
\end{align}

Collecting the homogeneous coefficients into  $\ph(\omega)=A_0+\omega A_1+\omega^2A_2+\cdots,$
$\mathcal M(\omega)=B_0+\omega B_1+\omega^2B_2+\cdots,$
and restoring the branch label yields
\begin{equation}
\Phi_\omega(r)\Big|_j=\ph_j(\omega)P_\ell(x)+\mathcal M_j(\omega)Q_\ell(x)+\delta\Phi^{\rm NZ}_{j,\omega}(r),
\qquad
\delta\Phi^{\rm NZ}_{j,\omega}=\omega^2\chi_j^{[2]}+{\cal O}(\omega^3).
\label{eq:bcS}
\end{equation}
In the overlap region, the field produced by distant external sources has a regular multipole expansion proportional to $r^\ell$, whereas an induced multipole produces a field proportional to $r^{-\ell-1}$. Since $P_\ell(x)$ and $Q_\ell(x)$ have these respective asymptotic behaviours \eqref{eq:legasym}, $\ph_j$ specifies the applied tidal amplitude in the chosen basis normalization,  while $\mathcal M_j$ is the coefficient of the induced decaying field. The basis functions are static, but their coefficients retain the frequency dependence imposed by the boundary conditions, while the inhomogeneous deformation must be retained when computing the second-order response and its logarithms.

Evaluating \eqref{eq:bcS} at the matching surface gives
\begin{equation}
\varphi_j(\omega)=\ph_j(\omega)P_\ell(x_R)+\mathcal M_j(\omega)Q_\ell(x_R)
+\delta\Phi^{\rm NZ}_{j,\omega}(R), \qquad x_R:=\frac{2R}{\rs}-1.
\end{equation}
The Dirichlet source $\varphi_j$ is the total field at the matching surface, whereas the tidal source $\phi_j$ is the coefficient of the growing mode. Similarly, $\mathcal M_j$ is the coefficient of the decaying mode, rather than directly
the expectation value of the induced multipole operator. The normalization relating the two is determined by varying the renormalized boundary action with respect to the tidal source. Thus we have decomposed the low-frequency field into growing and decaying components, together with near-zone corrections beginning at second order in frequency.

\subsection{Tidal response and leading absorption}
\label{sub:tidalresponse}
\noindent
The leading response can be obtained without the explicit MST series. The required ingredients are frequency reflection, complex conjugation, the exact horizon Wronskian \eqref{eq:Whor}, and the static Wronskian \eqref{eq:WPQ1}.  To extract the tidal response, we apply the near-zone decomposition \eqref{eq:bcS} to the horizon-normalized solutions $\psi_\pm$. In the overlap region \eqref{eq:region} we write
\begin{equation}
\psi_\pm(r,\omega)=\mathcal S_\pm(\omega)P_\ell(x)+\mathcal R_\pm(\omega)Q_\ell(x)+\delta\psi_\pm^{\rm NZ}(r,\omega).
\label{eq:connection}
\end{equation}
Here $\mathcal S_\pm$ and $\mathcal R_\pm$ are the growing and decaying coefficients of each horizon-normalized solution, while $\delta\psi_\pm^{\rm NZ}$ contains the inhomogeneous radial corrections beginning at order $\omega^2$. The labels $\pm$ distinguish outgoing and ingoing solutions.

For a physical ingoing field $\Phi_\omega=c(\omega)\psi_-$, comparison with \eqref{eq:bcS} gives $\ph=c\mathcal S_-$ and $\mathcal M=c\mathcal R_-$. We therefore define the radial response ratios, independent of the overall normalization of $\psi_\pm$,
\begin{equation}
\varrho_\pm(\omega):=\frac{\mathcal R_\pm(\omega)}{\mathcal S_\pm(\omega)}, \quad\mbox{with}\quad  \varrho_-(\omega)=\frac{\mathcal M}{\ph}, \label{eq:rhoDef}
\end{equation}
thus $\varrho_-$ measures the decaying-field coefficient per unit growing-field coefficient for the ingoing solution.
With a real matching prescription respecting
$\omega\to-\omega$, \eqref{eq:crossing} implies  $\varrho_+(\omega)=\varrho_-(-\omega)=\varrho_-(\omega)^*.$ At zero frequency, the horizon-regular solution with unit horizon value is $P_\ell$, so
\begin{equation}
\mathcal S_-(0)=1, \qquad \mathcal R_-(0)=0, \qquad \varrho_-(0)=0.
\label{eq:static0}
\end{equation}
The absence of a decaying component is the vanishing of the static scalar Love response.

Although $\varrho_-$ characterizes the radial response, its normalization relative to the tidal kernel is fixed by the
boundary action \eqref{eq:WbdyS}. At zero frequency, the solution $\psi_-=P_\ell$ is purely
growing, but its Dirichlet kernel is generally nonzero;  \eqref{eq:KpmExact}  gives
\begin{equation}
\mathcal K_-(R,0)=-f(R)R^2\left.\frac{\partial_rP_\ell(x)}{P_\ell(x)}\right|_{r=R},
\end{equation}
the boundary response of the pure static growing field. For each fixed multipole, we adopt the time-local static reference subtraction $K_{\ell,\rm ct}(R,\omega)=-\mathcal K_-(R,0)+{\cal O}(\omega^2)$ in \eqref{eq:KExactCutoff}. Consequently,
\begin{equation}
K_\ell^R(\omega)=\mathcal K_-(R,\omega)-\mathcal K_-(R,0)+{\cal O}(\omega^2).
\label{eq:klk0}
\end{equation}
We then convert from the Dirichlet source $\varphi$ to the
applied tidal amplitude $\ph$.

Using \eqref{eq:connection} and \eqref{eq:rhoDef}, we have $\psi_-=\mathcal S_-(P_\ell+\varrho_-Q_\ell)+{\cal O}(\omega^2)$. Since $\varrho_-={\cal O}(\omega)$, \eqref{eq:bcS} gives $\varphi=P_\ell(x_R)\ph+{\cal O}(\omega)$. The subtracted Dirichlet kernel is also ${\cal O}(\omega)$, so this leading source conversion suffices through first order in frequency. Substituting it into the quadratic boundary functional \eqref{eq:WbdyS} multiplies the kernel by $P_\ell(x_R)^2$. Using \eqref{eq:KpmExact} and the Wronskian \eqref{eq:WPQ1}, we evaluate the source-rescaled kernel in \eqref{eq:klk0}:
\begin{equation}
P_\ell^2 \left[ -fr^2\partial_r\log\psi_- +fr^2\frac{\partial_rP_\ell}{P_\ell} \right]_{r=R}
=\frac{\rs}{2}\, \varrho_-\, \frac{P_\ell(x_R)}{P_\ell(x_R)+\varrho_-Q_\ell(x_R)} +{\cal O}(\omega^2).
\end{equation}
The fraction equals $1+{\cal O}(\omega)$, and its correction therefore contributes only at ${\cal O}(\omega^2)$ after
multiplication by $\varrho_-$.  Thus
\begin{equation}
K_\ell^{\rm tidal}
=P_\ell(x_R)^2\bigl[\mathcal K_-(R,\omega)-\mathcal K_-(R,0)\bigr]+{\cal O}(\omega^2)
=\frac{\rs}{2}\,\varrho_-(\omega)+{\cal O}(\omega^2).
\label{eq:Kexact}
\end{equation}
The leading response normalization is therefore the weighted Wronskian \eqref{eq:WPQ1}, $-\mathcal W[P_\ell,Q_\ell]=\frac{\rs}{2}.$ At second and higher orders, the frequency-dependent near-zone deformation and matching counterterms must also be included. Equation \eqref{eq:Kexact} is therefore a leading low-frequency relation.

We now evaluate the leading absorptive term of the tidal kernel. The normalization relation \eqref{eq:Kexact} alone does not determine its coefficient; this follows from the horizon flux, or equivalently from the finite-radius absorption formula \eqref{eq:ImKcutoff}. For the local matched response at a fixed renormalization scale, we write
\begin{equation}
K_\ell^{\rm tidal}(\omega)=c_\ell^{(0)}+i\omega c_\ell^{(1)}
+\omega^2c_\ell^{(2)}+i\omega^3c_\ell^{(3)}+\cdots, \qquad c_\ell^{(0)}=0.
\label{eq:Kexp}
\end{equation}
The coefficients are real by
$K_\ell^{\rm tidal}(-\omega)=K_\ell^{\rm tidal}(\omega)^*$,
and $c_\ell^{(0)}=0$ expresses the vanishing static Love response. The conservative coefficients may depend on the renormalization scale; this local expansion excludes the additional long-distance frequency dependence of the full scattering response.

The static subtraction is real, while the conversion to tidal sources multiplies the absorptive kernel by $P_\ell(x_R)^2$. Taking $\omega\to0$ at fixed matching radius $R$ and using $\psi_-(R,0)=P_\ell(x_R)$ gives
\begin{equation}
c_\ell^{(1)}=\lim_{\omega\to0}\frac{{\rm Im}\,K_\ell^{\rm tidal}(\omega)}{\omega}
=\rs^2\lim_{\omega\to0}\frac{P_\ell(x_R)^2}{|\psi_-(R,\omega)|^2}
=\rs^2.
\label{eq:deltal}
\end{equation}
The positive sign agrees with \eqref{eq:absign} and the independence of $\ell$ reflects the source normalization $P_\ell(1)=1$; it does not imply that physical absorption cross sections are independent of the multipole. 

For the monopole, $\ph_{00}=\sqrt{4\pi}\,\ph_{\rm constant}$, so the dissipative coefficient for an angle-independent
source is
\begin{equation}
4\pi c_0^{(1)}=4\pi\rs^2=\AH.
\label{eq:sWaveArea}
\end{equation}
This is the horizon-area normalization underlying the low-frequency absorption cross section of a minimally coupled massless scalar \cite{Unruh:1976fm,Das:1996we}.  The conversion to sources normalized as the coefficients of $r^\ell$ is given in Appendix~\ref{sub:sourceNormalization}.

In this subsection, we have extracted the radial response from the decaying-to-growing coefficient ratio of the ingoing solution and related it to the tidal kernel through the boundary action, static reference subtraction and source conversion. The horizon flux then fixes the leading absorptive coefficient, while the horizon-regular static solution gives a vanishing static Love
response.

\subsection{The running relation}
\label{sub:645}
\noindent
The fluctuation--dissipation theorem relates equilibrium noise to absorption. A separate relation connects the leading absorptive response of a Schwarzschild black hole to the logarithmic running of its conservative dynamical Love coefficients \cite{Rodriguez:2026iot,Combaluzier:2025eoc,Caron-Huot:2025tlq}. We now derive this relation directly from the scalar radial
equation.

The leading absorptive coefficient is $c_\ell^{(1)}=\rs^2$, as established in \eqref{eq:deltal}. Our task is to determine the logarithmic coefficient $b_\ell$ in
\begin{equation}
c_\ell^{(2)}(\mu)=a_\ell-b_\ell\log(\mu\rs),
\label{eq:brule}
\end{equation}
where $a_\ell$ is a finite, scheme-dependent coefficient.

\subsubsection{First-order response and absorption}
\label{subsub:firstOrderAbsorption}
\noindent
To prepare the second-order calculation, we first discuss the low-frequency expansion of the ingoing solution,
\begin{equation}
\psi_-(r,\omega)
=P_\ell(x)+\omega u_{1,\ell}(r)
+\omega^2u_{2,\ell}(r)+\cdots.
\label{eq:runningExpansion}
\end{equation}
The zeroth-order normalization follows from \eqref{eq:static0}. Substitution into \eqref{eq:eomSomega} gives
\begin{equation}
\mathcal L u_{1,\ell}=0, \qquad \mathcal L u_{2,\ell}=-\frac{r^2}{f}P_\ell(x).
\label{eq:runningEquations}
\end{equation}
Since $u_{1,\ell}$ solves the homogeneous static equation,
it is a linear combination of $P_\ell$ and $Q_\ell$.
Matching the logarithm in $y^{-i\omega\rs}=1-i\omega\rs\log y+\cdots$ fixes the coefficient of $Q_\ell$. At first order, the ingoing horizon condition fixes
\begin{equation}
u_{1,\ell}=s_{1,\ell}P_\ell+2i\rs Q_\ell,
\end{equation}
where $s_{1,\ell}$ is purely imaginary. The coefficient of $Q_\ell$ gives $\varrho_-=2i\omega\rs+{\cal O}(\omega^2)$.
Together with \eqref{eq:Kexact}, this reproduces $c_\ell^{(1)}=\rs^2$, in agreement with \eqref{eq:deltal}.

\subsubsection{Second-order response and logarithmic running}
\label{subsub:secondOrderRunning}
\noindent
At second order, the radial equation is inhomogeneous. To extract its logarithmic contribution, we write $u_{2,\ell}=U_\ell P_\ell+V_\ell Q_\ell$. The two coefficient functions introduce a redundancy in this representation, which we fix by imposing the standard variation-of-parameters constraint $U_\ell'P_\ell+V_\ell'Q_\ell=0$,  where primes denote radial derivatives. Substituting this decomposition into the inhomogeneous equation, and using $\mathcal LP_\ell=\mathcal LQ_\ell=0$ together with the constraint and \eqref{eq:WPQ1}, we obtain
\begin{equation}
V_\ell'=\frac{2}{\rs}\frac{r^3}{r-\rs}P_\ell\!\left(\frac{2r}{\rs}-1\right)^2.
\label{eq:runningVprime}
\end{equation}
For integer $\ell$, the numerator is a polynomial whose value at $r=\rs$ is $\rs^3$. Hence
\begin{equation}
r^3P_\ell\!\left(\frac{2r}{\rs}-1\right)^2=(r-\rs)\mathcal P_\ell(r)+\rs^3,
\label{eq:runningPolynomial}
\end{equation}
with $\mathcal P_\ell$ polynomial, and therefore by integrating \eqref{eq:runningVprime} we obtain
\begin{equation}
V_\ell(r) = \frac{2}{\rs}\int^r\dd r'\,\mathcal P_\ell(r')+2\rs^2\log\frac{r-\rs}{\rs}+C_\ell.
\label{eq:runningV}
\end{equation}
Direct substitution of $u_{2,\ell}$, using the constraint, gives $\mathcal W[P_\ell,u_{2,\ell}]=V_\ell\mathcal W[P_\ell,Q_\ell]$ and $\mathcal W[u_{2,\ell},Q_\ell]=U_\ell\mathcal W[P_\ell,Q_\ell]$. The coefficients of $P_\ell$ and $Q_\ell$ in $\psi_-$ at radius $r$ are therefore $\mathcal S_-=1+\omega s_{1,\ell}+\omega^2U_\ell$ and $\mathcal R_-=2i\omega\rs+\omega^2V_\ell$, and by \eqref{eq:rhoDef} their ratio at $r=R$ is
\begin{equation}\label{eq:rhow2}
\varrho_-(\omega;R)= 2i\omega\rs +\omega^2\left[V_\ell(R)-2i\rs s_{1,\ell}\right] +{\cal O}(\omega^3).
\end{equation}
Here the near-zone deformation $\delta\psi_-^{\rm NZ}$ of \eqref{eq:connection} has been absorbed into the $R$-dependent coefficients $U_\ell(R)$ and $V_\ell(R)$.   The source-normalization term $s_{1,\ell}$ contributes
only an $R$-independent second-order term, while $U_\ell(R)$ first enters the ratio at order $\omega^3$. 
The logarithmic dependence at second order therefore comes from $V_\ell(R)$. To determine the running of $K_\ell^{\rm tidal}$, we must convert this radial ratio to the source-conjugate kernel. Since \eqref{eq:Kexact} was established only through first order in frequency, we must check whether this conversion introduces an additional logarithm at second order.

For this purpose, we denote the statically subtracted and source-rescaled Dirichlet kernel appearing in \eqref{eq:Kexact} by 
$\widetilde K_\ell^R :=P_\ell(x_R)^2 \left(\mathcal K_-(R,\omega)-\mathcal K_-(R,0)\right).$ 
Using the local variation-of-parameters decomposition at $R$, the constraint and the Wronskian \eqref{eq:WPQ1} give
\begin{align}
\widetilde K_\ell^R=\frac{\rs}{2}\,
\frac{\varrho_-}{1+\varrho_-Q_\ell(x_R)/P_\ell(x_R)}=\frac{\rs}{2}\varrho_-
+2\omega^2\rs^3\frac{Q_\ell(x_R)}{P_\ell(x_R)}+{\cal O}(\omega^3),
\label{eq:runningNormalization}
\end{align}
where the second equation follows from \eqref{eq:rhow2}.  Since $Q_\ell/P_\ell$ has a decaying power-series expansion at large radius, this correction contains no $\log(R/\rs)$ term and does not affect the logarithmic running.

Combining \eqref{eq:runningNormalization}, \eqref{eq:runningV} and \eqref{eq:rhow2}, and separating the power-dependent matching terms in the overlap region \eqref{eq:region}, yields the logarithmic contribution
\begin{equation}
\widetilde K_\ell^R(\omega;R)\Big|_{\log}=\omega^2\rs^3\log\frac{R}{\rs}.
\label{eq:scalarRunningLog}
\end{equation}
To remove the dependence on the arbitrary matching radius $R$ from the local tidal coefficient, we subtract $\omega^2\rs^3\log(\mu R)$. Here $\mu$ is an arbitrary inverse-length renormalization scale defining the separation between the local tidal coefficient and the exterior propagation contribution. Since $\log(R/\rs)-\log(\mu R)=-\log(\mu\rs)$, the remaining renormalized local contribution is $-\omega^2\rs^3\log(\mu\rs)$. Comparison with \eqref{eq:brule}, together with \eqref{eq:deltal}, then yields
\begin{equation}
b_\ell=\rs^3=\rs\,c_\ell^{(1)}.
\label{eq:bruleproved}
\end{equation}
The unit coefficient has a geometric origin. Since $P_\ell$ is polynomial, the logarithm arises solely from the horizon
pole. The factor $2/\rs$ in \eqref{eq:runningVprime} cancels against the normalization $\rs/2$ in
\eqref{eq:runningNormalization}, giving
\begin{equation}
b_\ell=\mathop{\rm Res}_{r=\rs}\frac{r^2P_\ell^2}{f}=\frac{c_\ell^{(1)}}{f'(\rs)}.
\label{eq:runningResidue}
\end{equation}
The horizon slope $f'(\rs)=1/\rs$ thus fixes the relative coefficient, so the factor $\beta/4\pi$ in \eqref{eq:645beta} enters through the surface gravity of the radial operator. Both $c_\ell^{(1)}$ and $b_\ell$ are proportional to $P_\ell(1)^2$, the square of the horizon amplitude of the static solution, which therefore cancels from their ratio; a rescaling of the source multiplies both equally.  The result agrees with the generic-multipole calculation of Ref.~\cite{Combaluzier:2025eoc}, where only the constant term of $P_\ell$ in powers of $x-1$ generates the logarithm (see also Refs.~\cite{Charalambous:2021mea,Perry:2023wmm}), and the residue form is a special case of the general residue formula of Ref.~\cite{Garcia-Saenz:2026pcb}.

Equivalently, using \eqref{eq:Kexp} and
$\rs=\beta/(4 \pi)$,
\begin{equation}
-\mu\frac{\partial}{\partial\mu} K_\ell^{\rm tidal}(\omega;\mu)= \frac{\beta\omega}{4\pi}\, {\rm Im}\,K_\ell^{\rm tidal}(\omega;\mu)+{\cal O}(\omega^3).
\label{eq:645beta}
\end{equation}
This relation holds through second order in frequency; it is not an exact finite-frequency identity. The numerical coefficients in \eqref{eq:bruleproved} use the source multiplying $P_\ell$. At fixed source normalization, finite, $\mu$-independent conservative local counterterms can shift $a_\ell$ without changing the leading logarithmic coefficient or absorption. The scale dependence of the matched coefficient cancels against the complementary propagation contribution in physical observables.

Our scalar result is consistent with the leading expansion of the resummed response in \cite{Solon:2026ubm}. Since the leading dissipation--running relation enters the absorptive part of that construction, this agreement is a consistency check. For the monopole, the angle-integrated coefficients also agree with the scalar Raman-scattering matching of \cite{Ivanov:2024sds}; see Appendix~\ref{sub:mst}.

In this subsection, we solved the second-order radial equation and isolated its logarithmic contribution. We verified that converting the radial response to the tidal kernel introduces no additional large-radius logarithm, then renormalized the matching-radius dependence to obtain the conservative running. Comparing its coefficient with the leading absorptive coefficient showed that their ratio is fixed by the Schwarzschild horizon slope, independently of thermal KMS symmetry.

\subsection{Relation to the fluctuation--dissipation theorem}
\label{sub:sameas}
\noindent
Having derived the running relation from the radial dynamics, we now compare it with the equilibrium fluctuation--dissipation theorem. The latter was established for the finite-radius Dirichlet kernels in \eqref{eq:fdtexact}; we first explain how it carries over to the matched tidal kernels.

At leading order, the source conversion used in \eqref{eq:Kexact} multiplies the response and noise kernels by the same real factor. The conservative local subtractions change neither the absorptive part nor the noise kernel. The tidal kernels therefore retain the fluctuation--dissipation relation,
\begin{equation}
F_\ell^{\rm tidal}(\omega)=\coth\!\left(\frac{\beta\omega}{2}\right)
\operatorname{Im}K_\ell^{\rm tidal}(\omega).
\label{eq:fdttidal}
\end{equation}
At higher orders, matching also involves the decaying component and frequency-dependent near-zone deformations, and must be
performed for the full equilibrium Schwinger--Keldysh functional. Preserving its KMS symmetry ensures that the matched kernels
obey the fluctuation--dissipation relation order by order. 

Combining the leading absorptive coefficient \eqref{eq:deltal}
with \eqref{eq:fdttidal} gives
\begin{equation}
F_\ell^{\rm tidal}(0)
=\lim_{\omega\to0}
\coth\!\left(\frac{\beta\omega}{2}\right)
{\rm Im}\,K_\ell^{\rm tidal}(\omega)
=\frac{2c_\ell^{(1)}}{\beta}
=\frac{2\rs^2}{\beta}
=\frac{\rs}{2\pi}.
\label{eq:Fzero}
\end{equation}
This result uses the source multiplying $P_\ell$ and the harmonic normalization \eqref{eq:Ynorm}. It also follows directly from the finite-radius result: as $\omega\to0$, \eqref{eq:ImKcutoff} with $\psi_-(R,0)=P_\ell(x_R)$ gives $F_\ell^R(0)=2\rs^2/[\beta P_\ell(x_R)^2]$, which reduces to \eqref{eq:Fzero} after the rescaling by $P_\ell(x_R)^2$.

The independent radial calculation established
$b_\ell=\rs c_\ell^{(1)}$ in \eqref{eq:bruleproved}.
Eliminating $c_\ell^{(1)}$ between the two relations yields
\begin{equation}
b_\ell=\frac{\beta\rs}{2}\,F_\ell^{\rm tidal}(0),
\label{eq:runningNoise}
\end{equation}
for the Hartle--Hawking state. For the scalar field considered here, the retarded kernel, and hence $b_\ell$ and $c_\ell^{(1)}$, are independent of the state, whereas the noise is not: in a state without thermal occupation, such as the Boulware state, the zero-frequency noise vanishes while the running relation \eqref{eq:645beta} is unchanged. A common real, frequency- and scale-independent source rescaling multiplies both sides by the same factor, preserving this relation.

Thus equilibrium noise and conservative logarithmic running are controlled by the same leading absorptive coefficient, but require different additional information. KMS analyticity fixes the noise in the Hartle--Hawking state, whereas the Schwarzschild radial dynamics and tidal matching fix the running. Equation \eqref{eq:runningNoise} combines these two results. KMS symmetry alone does not determine the conservative running, as also illustrated by \eqref{eq:kmsexp}.

\section{Conclusions}
\label{sec:disc}
\noindent
We have studied the real-time scalar response of a Schwarzschild black hole in the Hartle--Hawking state, using a minimally coupled
massless field and the Schwinger--Keldysh contour of Fig.~\ref{fig:contour}. The construction separates the finite-frequency boundary response from the low-frequency matching that defines the tidal Love coefficients.

The fluctuation--dissipation relation \eqref{eq:fdtexact} follows from the analytic continuation appropriate to the thermal state. The radial monodromy and the accompanying shift of Schwarzschild time produce the Bose--Einstein factor relating the two branches. The chosen Hartle--Hawking contour therefore implements the KMS condition, without an additional horizon matching condition. For the finite-radius Dirichlet problem, this derivation holds for every scalar multipole and arbitrary real frequency within the quadratic probe field problem. For the tidal kernels, the relation is preserved order by order by low-frequency matching consistent with KMS symmetry.

We have also derived the leading running relation \eqref{eq:645beta} directly from the scalar radial equation. The ingoing boundary condition fixes absorption through the horizon flux, while the second-order conservative logarithm is determined by the horizon residue
\eqref{eq:runningResidue}. The polynomial structure of the static Schwarzschild solution relates this residue to the matching logarithm.
The normalization of the static solution cancels from the ratio, and the Schwarzschild horizon slope fixes the unit coefficient in \eqref{eq:bruleproved}, independently of source normalization. This agrees with Ref.~\cite{Combaluzier:2025eoc}, while the comparison with the resummed response in Ref.~\cite{Solon:2026ubm} provides a consistency check; our derivation does not require its higher-order structure.

The two relations constrain different quantities. The fluctuation--dissipation theorem fixes equilibrium noise in terms of absorption, whereas the running relation requires the radial dynamics and tidal matching. In particular, quadratic KMS symmetry does not constrain the conservative coefficient at second order in frequency, as illustrated by \eqref{eq:kmsexp}. In the present  model, the retarded kernel is state-independent, so the running relation should also hold in the Boulware vacuum. The symmetric noise is state-dependent, and the thermal fluctuation--dissipation relation derived here
uses the KMS property of the Hartle--Hawking state.  Their common dependence on the leading absorptive response nevertheless gives \eqref{eq:runningNoise}, relating the logarithmic running to the zero-frequency thermal noise in the Hartle--Hawking state. This provides a precise connection between the two statements while keeping their distinct physical inputs explicit.

The contour derivation should extend to other static black holes admitting a regular equilibrium state and the corresponding thermal continuation. The tidal coefficients and their running, however, require a separate matching calculation in each background. Gravitational perturbations provide a natural next application, with the source normalization and radial dynamics treated separately in each parity sector.

For rotating black holes, the horizon frequency includes the angular chemical potential. Extending the construction to Kerr therefore requires an explicit treatment of superradiance and the quantum state, as an unrestricted asymptotically flat bosonic Hartle--Hawking equilibrium cannot be assumed~\cite{Kay:1988mu}. A suitable choice of state and boundary conditions would provide the starting point for comparing thermal fluctuations with the running of the rotating tidal response.

\begin{acknowledgments}
\noindent
We thank Denis Comelli, Valerio De Luca,  and Davide Perrone for useful discussions. D.G.  acknowledges support from  the National Science and Technology Council (NSTC) of Taiwan with the Young Scholar Columbus Fellowship grant 114-2636-M-110-004 and 115-2112-M-110-010.  A.R.  acknowledges support from the Swiss National Science Foundation (project number CRSII5\_213497). Large language models (Claude and ChatGPT) were used for editorial assistance and language polishing.
\end{acknowledgments}

\appendix

\section{Dynamical KMS symmetry}
\label{app:kms}
\noindent
This appendix verifies that the quadratic action obtained from the thermal contour satisfies dynamical KMS symmetry.
For the time-reversal-even scalar source in the equilibrium Schwarzschild problem, the transformation is
\begin{equation}
\widetilde\varphi_1(t) = \varphi_1(-t+i\theta),
\qquad \widetilde\varphi_2(t) = \varphi_2\bigl(-t-i(\beta-\theta)\bigr), \qquad 0\leq\theta\leq\beta,
\label{eq:kms}
\end{equation}
with the invariance condition $W[\widetilde\varphi_1,\widetilde\varphi_2]= W[\varphi_1,\varphi_2]$
\cite{Crossley:2015evo,Liu:2018kfw}. Here $\varphi_j$ are the finite-radius sources used in Sec.~\ref{sec:fdtexact}.

By time-translation invariance the condition does not depend on $\theta$; we take $\theta=\beta/2$ and introduce $\boldsymbol{\varphi}=(\varphi_r,\varphi_a)^T$. Using the Fourier convention \eqref{eq:FourierConvention}, the transformation becomes $\widetilde{\boldsymbol{\varphi}}(\omega) =S(\omega)\boldsymbol{\varphi}(-\omega)$, where
\begin{equation}
S(\omega) = \begin{pmatrix}\cosh\frac{\beta\omega}{2}
&-\frac12\sinh\frac{\beta\omega}{2}
\\
-2\sinh\frac{\beta\omega}{2}
&\cosh\frac{\beta\omega}{2}
\end{pmatrix}.
\label{eq:KMSmatrixS}
\end{equation}
The quadratic action \eqref{eq:WexactRA} can be written as
\begin{equation}
W=\frac12\int\frac{\dd\omega}{2\pi}\, \boldsymbol{\varphi}(-\omega)^T \mathbb M(\omega)\boldsymbol{\varphi}(\omega), \qquad
\mathbb M(\omega)=
\begin{pmatrix}
0 & K_\ell^R(-\omega)\\
K_\ell^R(\omega) & iF_\ell^R(\omega)
\end{pmatrix}.
\label{eq:kmsmatrix}
\end{equation}
Transforming the sources and relabelling the frequency gives the invariance condition
\begin{equation}
S(\omega)^T \mathbb M(-\omega) S(-\omega)=\mathbb M(\omega).
\label{eq:KMSinvariance}
\end{equation}

To see its content explicitly, we use $K_\ell^R(-\omega)=K_\ell^R(\omega)^*$ and $F_\ell^R(-\omega)=F_\ell^R(\omega)$; then the $rr$ component of \eqref{eq:KMSinvariance} is
\begin{equation}
4i \sinh\left(\frac{\beta\omega}{2}\right)\left[\cosh\left(\frac{\beta\omega}{2}\right)\,{\rm Im}\,K_\ell^R(\omega)-\sinh\left(\frac{\beta\omega}{2}\right)\,F_\ell^R(\omega)\right]=0.
\end{equation}
For nonzero real frequency, this is precisely \eqref{eq:fdtexact}; the zero-frequency relation follows by continuity.  Substituting it into the $ar$ component gives
\begin{equation}\cosh^2\left(\frac{\beta\omega}{2}\right)K_\ell^R(\omega)^* + 2i \cosh^2\left(\frac{\beta\omega}{2}\right)\, {\rm Im}\,K_\ell^R(\omega) - \sinh^2\left(\frac{\beta\omega}{2}\right) K_\ell^R(\omega)=K_\ell^R(\omega)
\end{equation}
as required; the $ra$ and $aa$ components follow in the same way.

The derivative counting is also explicit in the low-frequency expansion. For the matched tidal kernel \eqref{eq:Kexp}, the fluctuation--dissipation relation gives
\begin{equation}
F_\ell^{\rm tidal}(\omega)=\frac{2c_\ell^{(1)}}{\beta}+\omega^2\left(\frac{\beta c_\ell^{(1)}}{6}+\frac{2c_\ell^{(3)}}{\beta}\right)+{\cal O}(\omega^4).
\label{eq:kmsexp}
\end{equation}
Thus the noise at second order in frequency requires the dissipative response at third order. At quadratic order,
dynamical KMS invariance is equivalent to the fluctuation--dissipation relation and places no additional constraint on the real, even-in-frequency response. In particular, the running of $c_\ell^{(2)}$ requires the separate radial calculation of Sec.~\ref{sec:running}.

\section{Comparison with the resummed response}
\label{sub:mst}
\noindent
The same leading relation is contained in the factorized response of \cite{Solon:2026ubm},
\begin{equation}
\frac{\bar{\mathcal F}_{\ell,s}}{4\pi\rs^{2\ell+1}}=G_{\ell,s}(\bar y)-\frac12\eta\,G_{\ell,s}'(\bar y),
\qquad \bar y=-\frac12\eta^2{\mathcal T},\qquad
{\mathcal T}=\log\frac{\rs}{R}-2\sum_{k\ge2}\zeta_k\eta^{k-1}.
\label{eq:Fclosed}
\end{equation}
Here $\eta$ is defined in \eqref{eq:etaIntro}, $\zeta_k=\zeta(k)$, $s$ denotes the perturbation spin, and the prime denotes
differentiation with respect to the argument of $G_{\ell,s}$. This expression is a result imported from the cited resummation, rather than an additional consequence of the thermal contour. Its ingredients include horizon matching, near-zone evolution, and long-distance dressing \cite{Kosmopoulos:2025rfj,Solon:2026ubm}. Although the cited expression applies to several perturbation spins, its comparison with our calculation concerns $s=0$.

For vanishing static Love response, $G_{\ell,s}(0)=0$. Writing $L_{\ell,s}^{(1)}=G_{\ell,s}'(0)$ gives
\begin{equation}
\frac{\bar{\mathcal F}_{\ell,s}}{4\pi\rs^{2\ell+1}} =-\frac12L_{\ell,s}^{(1)}\left[\eta+\eta^2\log\frac{\rs}{R}\right]+{\cal O}(\eta^3).
\label{eq:FclosedExpand}
\end{equation}
The normalization of $\bar{\mathcal F}_{\ell,s}$ differs from that of our source-conjugate kernel. For the two terms displayed, a real, frequency-independent conversion factor multiplies both coefficients and cancels from their ratio.

Substituting $\eta=i\omega\rs$ into \eqref{eq:FclosedExpand} gives
\begin{equation}
\frac{\bar{\mathcal F}_{\ell,s}}{4\pi\rs^{2\ell+1}}=-\frac12L_{\ell,s}^{(1)}\left[i\omega\rs+\omega^2\rs^2\log\frac{R}{\rs}\right]+{\cal O}(\omega^3).
\end{equation}
After conversion to our source normalization, the coefficient of $i\omega$ is $c_\ell^{(1)}$, so the logarithmic term is
$\omega^2\rs c_\ell^{(1)}\log(R/\rs)$. After the logarithmic subtraction described below \eqref{eq:scalarRunningLog}, this becomes $-\omega^2\rs c_\ell^{(1)}\log(\mu\rs)$. Comparison with \eqref{eq:brule} therefore gives
\begin{equation}
b_\ell=\rs\,c_\ell^{(1)},
\label{eq:bruleMST}
\end{equation}
in agreement with \eqref{eq:bruleproved}. The far-zone dressing in ${\mathcal T}$ first affects higher
orders in this expansion. Those orders involve additional dynamical information; the leading relation \eqref{eq:645beta} should not be
iterated unchanged.

For the monopole, the leading coefficients can also be compared with the scalar Raman-scattering matching of \cite{Ivanov:2024sds},
performed in dimensional regularization with  modified minimal subtraction. Summing over the sphere as in \eqref{eq:sWaveArea}, our results give the angle-integrated dissipative coefficient $4\pi c_0^{(1)}=\AH$ and logarithmic coefficient
$4\pi b_0=\AH\,\rs$. These agree with the leading absorptive coefficient and the  modified minimal subtraction running obtained there. For absorption, the comparison is made at positive frequency, accounting for the time-ordered correlator used in that calculation and the retarded convention used here. We compare the leading absorptive and logarithmic conservative coefficients; the finite
conservative term $a_0$ depends on the subtraction prescription.

\section{Source normalization and absorption cross sections}
\label{sub:sourceNormalization}
\noindent
For the monopole, a spatially constant source satisfies $\ph_{00}=\sqrt{4\pi}\,\ph_{\rm constant}$. Its angle-integrated dissipative coefficient is given in \eqref{eq:sWaveArea}, and is the horizon-area normalization underlying the low-frequency absorption cross section of a minimally coupled massless scalar \cite{Unruh:1976fm,Das:1996we}.

For a general multipole, suppose the growing radial field is written as $\mathcal E\,r^\ell$. The large-radius behaviour
\begin{equation}
P_\ell\!\left(\frac{2r}{\rs}-1\right)\sim \frac{(2\ell)!}{(\ell!)^2\rs^\ell}\,r^\ell
\end{equation}
then gives
$\ph=\alpha_\ell\mathcal E$, where $\alpha_\ell\equiv(\ell!)^2\rs^\ell/(2\ell)!$. Substitution into the response term of \eqref{eq:Wdef}
yields
\begin{equation}
\ph_a(-\omega)K_\ell^{\rm tidal}(\omega)\ph_r(\omega)=\mathcal E_a(-\omega) K_\ell^{(\mathcal E)}(\omega)\mathcal E_r(\omega),
\qquad
K_\ell^{(\mathcal E)}(\omega)= \alpha_\ell^2K_\ell^{\rm tidal}(\omega).
\end{equation}
The leading dissipative coefficient in this normalization is therefore
\begin{equation}
\hat c_\ell^{(1)} \equiv \lim_{\omega\to0} \frac{{\rm Im}\,K_\ell^{(\mathcal E)}(\omega)}{\omega}=\rs^{2\ell+2}\frac{(\ell!)^4}{[(2\ell)!]^2}.
\label{eq:c1ell}
\end{equation}
This coefficient is defined using the same normalized harmonics as \eqref{eq:Ynorm}. If the decaying radial term is $\mathcal D\,r^{-\ell-1}$, then at leading order $K_\ell^{(\mathcal E)}=(2\ell+1)\mathcal D/\mathcal E$.
Thus a radial response-to-source ratio and a source-conjugate kernel carry different normalization
factors. Comparison with a worldline EFT convention additionally requires the corresponding multipole-operator normalization. The linear-frequency tidal response does not imply multipole-independent absorption cross sections. For fixed incident wave amplitude at infinity, the near-zone tidal amplitude scales as $\omega^\ell$, giving $\sigma_\ell\propto\omega^{2\ell}$ in the low-frequency limit.

\bibliographystyle{JHEP}
\bibliography{refs3}

\end{document}